\documentclass[pra,aps,onecolumn,superscriptaddress,showkeys,nofootinbib]{revtex4}
\usepackage[latin1]{inputenc}
\usepackage{amsmath,amsthm,amssymb,graphicx,subfigure,xcolor}
\usepackage{mathrsfs}
\usepackage{graphics,float}
\usepackage{natbib}
\usepackage{graphicx}
\usepackage{epstopdf}
\usepackage{color}
\usepackage[unicode=true]{hyperref}
\usepackage{booktabs}
\usepackage{tabularx}
\usepackage{tabu}
\usepackage{multirow}
\allowdisplaybreaks[4]
\hypersetup{
     colorlinks=true,       		
     linkcolor=blue,          	
     citecolor=blue,            
     urlcolor=blue,           	
 }

\newtheorem*{theorem*}{Theorem}

\newtheorem*{corollary*}{Corollary}

\newtheorem*{lemma*}{Lemma}

\newtheorem*{proposition*}{Proposition}
\theoremstyle{definition}

\newtheorem*{definition*}{Definition}
\theoremstyle{remark}

\newtheorem*{remark*}{Remark}

\begin{document}
\title{Tighter monogamy and polygamy relations in multiparty quantum systems}

\author{Chenxiao Wang}
\affiliation{School of Mathematics and Science, Hebei GEO University, Shijiazhuang 052161, China}

\author{Limin Gao}
\email{gaoliminabc@163.com}

\affiliation{School of Mathematics and Science, Hebei GEO University, Shijiazhuang 052161, China}
\affiliation{Intelligent Sensor Network Engineering Research Center of Hebei Province, Hebei GEO University, Shijiazhuang 052161, China}

\begin{abstract}
The monogamy and polygamy properties of quantum entanglement characterize fundamental constraints on the distribution of entanglement in multipartite quantum systems. In this paper, we investigate tighter monogamy and polygamy relations for multipartite entanglement. By establishing a new mathematical inequality, we derive a family of improved monogamy and polygamy inequalities for tripartite quantum systems and further extend these results to general multipartite systems. Comparisons with existing results show that the obtained bounds are tighter. Illustrative examples are provided to demonstrate the effectiveness of the proposed relations.
\end{abstract}



\maketitle
\section{Introduction}

Multipartite quantum entanglement plays a central role in quantum information processing tasks such as quantum cryptography, quantum communication, and quantum computation. One of the fundamental features of quantum entanglement is the monogamy of entanglement (MoE) \cite{12,13}, which characterizes the restrictions on how entanglement can be shared among multiple subsystems. In particular, such constraints are crucial for guaranteeing the security of many quantum information protocols \cite{3,4,5,6}. The quantitative characterization of entanglement monogamy is usually expressed in the form of inequalities. It was first formulated by Coffman, Kundu, and Wootters (CKW) for tripartite systems in terms of the squared concurrence \cite{7}, which is now known as the CKW inequality \cite{12}. Osborne and Verstraete subsequently generalized the CKW inequality from tripartite systems to $N$-qubit systems \cite{14}. Since then, the monogamy of entanglement for various entanglement measures has been extensively investigated \cite{16,17,18,19,20,21,22,23,24,25,26,27,28,29,30,31,32,33a,34}.

On the other hand, the polygamy of entanglement (POE) provides another perspective for understanding the distribution of entanglement in multipartite quantum systems. In particular, assisted entanglement measures, which can be regarded as dual quantities of bipartite entanglement measures, are known to satisfy polygamy relations. In 2007, Gour \textit{et al.} established the first polygamy inequality for three-qubit systems in terms of the squared concurrence of assistance (CoA) \cite{35}, which was later generalized to multiqubit systems \cite{36}. Subsequently, various assisted entanglement measures were shown to satisfy polygamy relations in multiqubit and higher-dimensional systems \cite{37a,37b,39,40}.

In the past decade, tighter monogamy and polygamy relations have attracted considerable attention \cite{41,33,42,43,44,37,45,46,47,49,50,51,52,53,54,55,56,57,58,59,60}. Although a number of improved inequalities have been proposed, obtaining tighter bounds for the distribution of multipartite entanglement remains an important problem.

In this paper, we investigate tighter monogamy and polygamy relations for multipartite quantum systems. By establishing a new mathematical inequality, we derive a family of improved monogamy and polygamy inequalities for tripartite quantum systems and further generalize these results to general multipartite systems. In addition, illustrative examples are presented to demonstrate that the obtained bounds are tighter than several existing results.

\section{Preliminaries}\label{a}

This section presents the notations, definitions, and basic inequalities that will be used throughout this paper. Consider an $N$-partite quantum state $\rho_{AB_1\cdots B_{N-1}}$ acting on the finite-dimensional Hilbert space
$\mathcal{H}_A \otimes \mathcal{H}_{B_1} \otimes \cdots \otimes \mathcal{H}_{B_{N-1}}$, with $\mathcal{S}$ denoting the set of all such states. We denote by $\rho_{A|B_1\cdots B_{N-1}}$ the state $\rho_{AB_1\cdots B_{N-1}}$ viewed as a bipartite state under the partition $A|B_1\cdots B_{N-1}$.

For a bipartite pure state $|\psi\rangle_{AB}=\sum_i \sqrt{\lambda_i}\,|ii\rangle$, the concurrence is defined as \cite{12,62}

\begin{equation}\label{120}
C(|\psi\rangle_{AB})=\sqrt{2\left[1-\operatorname{Tr}(\rho_A^2)\right]},
\end{equation}
where $\rho_A=\operatorname{Tr}_B(|\psi\rangle_{AB}\langle\psi|)$ is the reduced density matrix of subsystem $A$.

For a bipartite mixed state $\rho_{AB}$, the concurrence is defined through the convex-roof extension

\begin{equation}\label{s.j18}
C(\rho_{AB})=\min_{\{p_j,|\psi_j\rangle\}}\sum_j p_j C(|\psi_j\rangle_{AB}),
\end{equation}
where the minimum is taken over all pure-state decompositions of $\rho_{AB}$ satisfying
$\rho_{AB}=\sum_j p_j |\psi_j\rangle_{AB}\langle\psi_j|.$
For two-qubit states, the concurrence admits a closed analytical expression \cite{7}

\begin{equation}\label{3}
C(\rho)=\max\left\{0,\lambda_1-\lambda_2-\lambda_3-\lambda_4\right\},
\end{equation}
where $\lambda_i$ $(i=1,2,3,4)$ denote the eigenvalues, arranged in decreasing order, of
$X=\sqrt{\sqrt{\rho}(\sigma_y\otimes\sigma_y)\rho^*(\sigma_y\otimes\sigma_y)\sqrt{\rho}}.$

The concurrence of assistance (CoA) for a bipartite mixed state $\rho_{AB}$ is defined as \cite{63}

\begin{equation}\label{s.19}
C_a(\rho_{AB})=\max_{\{p_j,|\psi_j\rangle\}}\sum_j p_j C(|\psi_j\rangle_{AB}),
\end{equation}
where the maximum is taken over all pure-state decompositions of $\rho_{AB}$.

For any $N$-party quantum state
$\rho_{AB_1\cdots B_{N-1}} \in \mathcal{H}_A \otimes \mathcal{H}_{B_1} \otimes \cdots \otimes \mathcal{H}_{B_{N-1}}$,
let $E$ be an entanglement measure, and let $\alpha$ denote the infimum of all exponents such that $E^\alpha$ satisfies the monogamy relation
\begin{equation}\label{4}
E^\alpha(\rho_{A|B_1\cdots B_{N-1}})
\geq
\sum_{i=1}^{N-1} E^\alpha(\rho_{AB_i}).
\end{equation}
That is,
\begin{equation}\label{6}
\alpha
=
\inf \left\{
\eta :
E^\eta(\rho_{A|B_1\cdots B_{N-1}})
\geq
\sum_{i=1}^{N-1} E^\eta(\rho_{AB_i}),
\ \text{for all } \rho_{AB_1\cdots B_{N-1}} \in \mathcal{S}
\right\},
\end{equation}
where
$\rho_{AB_i}=\operatorname{Tr}_{\overline{B_i}}
\bigl(\rho_{AB_1\cdots B_{N-1}}\bigr),
\qquad i=1,2,\ldots,N-1,$ and $\overline{B_i}$ denotes the complement of $B_i$.

Similarly, for an entanglement of assistance measure $E_a$, let $\beta$ denote the supremum of all exponents such that $E_a^\beta$ satisfies the polygamy relation
\begin{equation}\label{5}
E_a^\beta(\rho_{A|B_1\cdots B_{N-1}})
\leq
\sum_{i=1}^{N-1} E_a^\beta(\rho_{AB_i}).
\end{equation}
That is,
\begin{equation}\label{88}
\beta
=
\sup \left\{
\gamma :
E_a^\gamma(\rho_{A|B_1\cdots B_{N-1}})
\leq
\sum_{i=1}^{N-1} E_a^\gamma(\rho_{AB_i}),
\ \text{for all } \rho_{AB_1\cdots B_{N-1}} \in \mathcal{S}
\right\}.
\end{equation}

\textbf{Lemma 1} \cite{33}. For $\mu,v \ge 0$, and $0\leq x\leq \frac{1}{k}$ with $k\geq1$. Then
\begin{equation}\label{s1.000}
\begin{aligned}
(1+x)^{\mu}\geq & 1+\frac{k \mu}{k+1} x+[(k+1)^{\mu}-(1+\frac{\mu}{k+1})k^{\mu}]x^{\mu}\\
\geq & 1+[(k+1)^{\mu}-k^{\mu}]x^{\mu}\geq 1+(2^{\mu}-1)x^{\mu}\\
\geq & 1+\mu x^{\mu}
\end{aligned}
\end{equation}
for $\mu\geq1$; and
\begin{equation}\label{00}
\begin{aligned}
(1+x)^{v}\leq & 1+\frac{k^2v}{(k+1)^2} x+\left((k+1)^{v}-\left[\frac{kv}{(k+1)^2}+1\right]k^{v}\right)x^{v}\\
\leq & 1+[(k+1)^{v}-k^{v}]x^{v}\leq1+(2^{v}-1)x^{v}\\
\leq & 1+v x^{v}
\end{aligned}
\end{equation}
for $0\leq v\leq1$.

\textbf{Lemma 2} \cite{37}. For \(0 \leq x \leq \frac{1}{k}\), \(k \geq 1\) and \(\mu \geq 2\). Then,
\begin{equation}\label{7}
(1 + x)^\mu \geq 1 + \mu x + \left[(k + 1)^\mu - \mu k^{\mu - 1} - k^\mu\right]x^\mu.
\end{equation}

\textbf{Lemma 3.}
For $0 \le x \le 1$ and $\mu \ge 3$, we have
\begin{equation}\label{8}
(1+x)^\mu \ge 1 + \mu x + \frac{\mu(\mu-1)}{2}x^2
+ \left(2^\mu - \frac{\mu(\mu+1)}{2} - 1\right)x^\mu .
\end{equation}

\textbf{Proof.}
The inequality is obvious when $x=0$. For $x>0$, define
$f(x,\mu)=\frac{(1+x)^\mu-1-\mu x-\frac{\mu(\mu-1)}{2}x^2}{x^\mu}.$
Differentiating $f$ with respect to $x$ yields
$f'(x,\mu)=\frac{\mu}{x^{\mu+1}}
\left[1+(\mu-1)x+\frac{(\mu-1)(\mu-2)}{2}x^2-(1+x)^{\mu-1}
\right].$
Let $g(x,\mu)=1+(\mu-1)x+\frac{(\mu-1)(\mu-2)}{2}x^2-(1+x)^{\mu-1}.$
Then
$\frac{\partial g}{\partial x}=(\mu-1)\left[1+(\mu-2)x-(1+x)^{\mu-2}\right].$
Since $\mu\ge3$ and $x\ge0$, Bernoulli's inequality gives
$(1+x)^{\mu-2} \ge 1+(\mu-2)x,$
which implies
$\frac{\partial g}{\partial x} \le 0.$
Hence $g(x,\mu)$ is decreasing with respect to $x$ on $[0,1]$. Moreover,
$g(x,\mu) \le g(0,\mu)=0.$
Therefore $f'(x,\mu)\le0$, and thus $f(x,\mu)$ is decreasing on $(0,1]$.
Consequently,
$f(x,\mu) \ge f(1,\mu)
=2^\mu-\mu-\frac{\mu(\mu-1)}{2}-1
=2^\mu-\frac{\mu(\mu+1)}{2}-1 .$
Multiplying both sides by $x^\mu$ yields
$(1+x)^\mu \ge 1+\mu x+\frac{\mu(\mu-1)}{2}x^2
+\left(2^\mu-\frac{\mu(\mu+1)}{2}-1\right)x^\mu .$
This completes the proof.
\qed

\textbf{Lemma 4.}
For $0 \le x \le \frac{1}{k}$, $k \ge 1$, and $\mu \ge 3$, we have
\begin{equation}\label{9}
(1+x)^\mu
\ge
1 + \mu x + \frac{\mu(\mu - 1)}{2}x^2
+
\left[(k+1)^\mu - \mu k^{\mu-1}
- \frac{\mu(\mu-1)}{2}k^{\mu-2} - k^\mu\right]x^\mu.
\end{equation}

\textbf{Proof.}
The inequality is trivial when $x=0$. For $x>0$, consider
$f(x,\mu)=\frac{(1+x)^\mu-1-\mu x-\frac{\mu(\mu-1)}{2}x^2}{x^\mu}.$
From the proof of Lemma 3, we know that $f(x,\mu)$ is decreasing with respect to $x$ on $(0,1]$ for $\mu\ge3$. Since $k\ge1$, we have $1/k\le1$, and therefore $f$ is decreasing on $(0,1/k]$.
Hence
$f(x,\mu) \ge f\!\left(\frac{1}{k},\mu\right).$
A direct computation gives
\[
f\!\left(\frac{1}{k},\mu\right)
=\frac{\left(1+\frac{1}{k}\right)^\mu-1-\frac{\mu}{k}-\frac{\mu(\mu-1)}{2k^2}}{(1/k)^\mu}
=(k+1)^\mu-\mu k^{\mu-1}-\frac{\mu(\mu-1)}{2}k^{\mu-2}-k^\mu .
\]

Thus
\[
f(x,\mu)
\ge
(k+1)^\mu-\mu k^{\mu-1}-\frac{\mu(\mu-1)}{2}k^{\mu-2}-k^\mu .
\]
Multiplying both sides by $x^\mu$ yields inequality (\ref{9}).
\qed

Note that, since $0\le kx\le 1$ and $\mu \ge 3$, we have
$(kx)^{\mu-2} \le 1.$ Hence
\begin{equation}\label{99}
\begin{split}
& 1 + \mu x + \frac{\mu(\mu - 1)}{2}x^2
+\left[(k+1)^\mu - \mu k^{\mu-1}- \frac{\mu(\mu-1)}{2}k^{\mu-2} - k^\mu\right]x^\mu\\
&=1 + \mu x + \frac{\mu(\mu - 1)}{2}x^2\left[1-(kx)^{\mu-2}\right]+\left[(k+1)^\mu - \mu k^{\mu-1} - k^\mu\right]x^\mu\\
&\ge 1 + \mu x + \left[(k+1)^\mu - \mu k^{\mu-1} - k^\mu\right]x^\mu\\
&\ge 1 + \left[(k+1)^\mu- k^\mu\right]x^\mu.
\end{split}
\end{equation}

\textbf{Lemma 5}. For \(0 \leq x \leq 1, 0 \leq v \leq 1\), we have
\begin{equation}\label{26}
(1+x)^v \leq 1+\frac{v}{2}x+\left(2^v-\frac{v}{2}-1\right)x^v.
\end{equation}

\textbf{Proof}. If $x=0$, the inequality is obvious. In the following, assume $0<x\le 1$.
Define $f(x,v)=\frac{(1+x)^v-\frac{v}{2}x-1}{x^v}.$
Then
 $\frac{\partial f}{\partial x}=\frac{v}{x^{v+1}}
\left[-(1+x)^{v-1}+\frac{v-1}{2}x+1\right].$
Let
$g(x,v)=-(1+x)^{v-1}+\frac{v-1}{2}x+1.$
A direct computation gives
 $ \frac{\partial^2 g}{\partial x^2}
=-(v-1)(v-2)(1+x)^{v-3}\le 0 $
for $0\le v\le 1$ and $0\le x\le 1$. Hence $g(x,v)$ is concave in $x$ on $[0,1]$, and its minimum on the interval $[0,1]$ is attained at one of the endpoints. Since
$g(0,v)=0$
and
$g(1,v)=1+\frac{v-1}{2}-2^{v-1}
=\frac{1+v-2^v}{2}\ge 0, $
it follows that $g(x,v)\ge 0$ for all $0\le x\le 1$. Thus,
$\frac{\partial f}{\partial x}\ge 0,$
which shows that $f(x,v)$ is increasing in $x$. Therefore,
 $f(x,v)\le f(1,v)=2^v-\frac{v}{2}-1.$
Multiplying both sides by $x^v$, we obtain
 $(1+x)^v \le 1+\frac{v}{2}x+\left(2^v-\frac{v}{2}-1\right)x^v.$
This completes the proof.
\qed

\textbf{Lemma 6}. For $0 \leq x \leq \frac{1}{k}$ with $k \geq 1$ and $0 \leq v \leq 1$, the following inequality holds
\begin{equation}\label{27}
(1+x)^v \leq 1 + \frac{kv}{k+1}x + \left[ (k+1)^v - \left(\frac{v}{k+1} + 1 \right) k^v\right]x^v
\end{equation}

\textbf{Proof}. The assertion is trivial for \(x=0\) or \(v=0\). We therefore assume
\(0<x\le 1/k\) and \(0<v\le 1\), and define
\[
f(x):=\frac{(1+x)^v-1-\frac{kv}{k+1}x}{x^v}.
\]
Then
\[
f'(x)=\frac{v}{x^{v+1}}\phi(x),
\qquad
\phi(x):=1+\frac{k(v-1)}{k+1}x-(1+x)^{v-1}.
\]
Moreover,
\[
\phi''(x)=-(v-1)(v-2)(1+x)^{v-3}\le 0,
\]
so \(\phi\) is concave on \([0,1/k]\). Since \(\phi(0)=0\), it remains to check \(\phi(1/k)\ge 0\). Writing \(s=1-v\in[0,1]\), we have
\[
\phi\!\left(\frac1k\right)
=1-\frac{s}{k+1}-\left(1-\frac1{k+1}\right)^s \ge 0,
\]
where the last inequality follows from
\[
(1-y)^s\le 1-sy,\qquad 0\le y\le 1,\quad 0\le s\le 1.
\]
Hence \(\phi(x)\ge 0\) on \([0,1/k]\), and therefore \(f'(x)\ge 0\) on \((0,1/k]\). Thus \(f\) is increasing, and
\[
f(x)\le f\!\left(\frac1k\right)
=\frac{\left(1+\frac1k\right)^v-1-\frac{v}{k+1}}{(1/k)^v}
=(k+1)^v-\left(1+\frac{v}{k+1}\right)k^v.
\]
Multiplying by \(x^v\) yields
\[
(1+x)^v \le 1+\frac{kv}{k+1}x
+\left[(k+1)^v-\left(1+\frac{v}{k+1}\right)k^v\right]x^v.
\]
This completes the proof.
\qed

Furthermore, since $0\le kx\le 1$ and $0\le v\le 1$, we have
$kx \le (kx)^v .$
Hence
$\frac{v}{k+1}[kx-(kx)^v]\le \frac{kv}{(k+1)^2}[kx-(kx)^v]\le 0.$
Consequently,
\begin{equation}\label{100}
\begin{split}
&1 + \frac{kv}{k+1}x
+ \left[ (k+1)^v - \left(\frac{v}{k+1} + 1 \right) k^v\right]x^v \\
&= 1 + \frac{v}{k+1}[kx-(kx)^v] + \big[(k+1)^v-k^v\big]x^v \\
&\le 1 + \frac{kv}{(k+1)^2}[kx-(kx)^v] + \big[(k+1)^v-k^v\big]x^v \\
&=1+\frac{k^2v}{(k+1)^2} x+\left((k+1)^{v}-\left[\frac{kv}{(k+1)^2}+1\right]k^{v}\right)x^{v} \\
&\leq 1 + \left[(k+1)^v - k^v \right]x^v.\\
\end{split}
\end{equation}

On the basis of these inequalities, our main results are presented in the following section.
\section{ Tighter monogamy relation}\label{b}

In this section, based on inequality (\ref{9}), we present several monogamy relations applicable to arbitrary quantum states. Moreover, the corresponding lower bounds are tighter.

\textbf{Theorem 1.}
Let $\rho_{AB_1B_2}\in \mathcal{H}_A \otimes \mathcal{H}_{B_1}\otimes \mathcal{H}_{B_2}$ be a tripartite quantum state and let $E$ be an entanglement measure. Suppose that $E^\alpha$ satisfies the monogamy relation. Let $\eta \ge 3\alpha$ and denote $\mu=\eta/\alpha\ (\mu\ge3)$.
For any $k\ge1$, the following results hold.

\textit{(i)} If $E^\alpha(\rho_{AB_1}) \ge k\,E^\alpha(\rho_{AB_2}),$
then
\begin{equation}\label{11}
\begin{split}
E^\eta(\rho_{A|B_1B_2}) \ge &
E^\eta(\rho_{AB_1})
+ \mu E^{\eta-\alpha}(\rho_{AB_1})E^\alpha(\rho_{AB_2})
+ \frac{\mu(\mu-1)}{2}E^{\eta-2\alpha}(\rho_{AB_1})E^{2\alpha}(\rho_{AB_2}) \\
&+\left[(k+1)^\mu
-\mu k^{\mu-1}
-\frac{\mu(\mu-1)}{2}k^{\mu-2}
-k^\mu\right]E^\eta(\rho_{AB_2}).
\end{split}
\end{equation}

\textit{(ii)} If $E^\alpha(\rho_{AB_2}) \ge k\,E^\alpha(\rho_{AB_1}),$
then
\begin{equation}\label{12}
\begin{split}
E^\eta(\rho_{A|B_1B_2}) \ge &
E^\eta(\rho_{AB_2})
+ \mu E^{\eta-\alpha}(\rho_{AB_2})E^\alpha(\rho_{AB_1})
+ \frac{\mu(\mu-1)}{2}E^{\eta-2\alpha}(\rho_{AB_2})E^{2\alpha}(\rho_{AB_1}) \\
&+\left[(k+1)^\mu
-\mu k^{\mu-1}
-\frac{\mu(\mu-1)}{2}k^{\mu-2}
-k^\mu\right]E^\eta(\rho_{AB_1}).
\end{split}
\end{equation}

\textbf{Proof.}
Since \(E^\alpha\) satisfies the monogamy relation, we have
\[
E^\alpha(\rho_{A|B_1B_2})
\ge E^\alpha(\rho_{AB_1})+E^\alpha(\rho_{AB_2}).
\]
Since $\eta=\mu\alpha$ and $\mu\ge 3$, it follows that
\[
E^\eta(\rho_{A|B_1B_2})
=\bigl(E^\alpha(\rho_{A|B_1B_2})\bigr)^\mu
\ge \bigl(E^\alpha(\rho_{AB_1})+E^\alpha(\rho_{AB_2})\bigr)^\mu.
\]
Therefore,
\[
E^\eta(\rho_{A|B_1B_2})
\ge E^\eta(\rho_{AB_1})
\left(1+\frac{E^\alpha(\rho_{AB_2})}{E^\alpha(\rho_{AB_1})}\right)^\mu.
\]
Moreover, from
$E^\alpha(\rho_{AB_1})\ge kE^\alpha(\rho_{AB_2}),$
we obtain
$0\le \frac{E^\alpha(\rho_{AB_2})}{E^\alpha(\rho_{AB_1})}\le \frac1k\le1.$
Applying inequality (\ref{9}) with
$x=\frac{E^\alpha(\rho_{AB_2})}{E^\alpha(\rho_{AB_1})},$
we obtain  (\ref{11}). The proof of (\ref{12}) is analogous. Interchanging the roles of \(B_1\) and \(B_2\) in the above argument yields (\ref{12}).
\qed

By inequality (\ref{99}), it follows that the lower bounds in Theorem 1 are greater than those in [\citealp{37}, Theorem 3].

In particular, when \( E^\alpha (\rho_{AB_1}) \geq E^\alpha (\rho_{AB_2}) \), $k=1$, $\mu\geq3$, we have
\begin{equation}\label{14}
\begin{split}
E^\eta(\rho_{A|B_1B_2})\geq & E^\eta(\rho_{AB_1}) + \mu E^{\eta-\alpha}(\rho_{AB_1}) E^\alpha(\rho_{AB_2}) +\frac{\mu(\mu - 1)}{2}E^{\eta - 2\alpha}(\rho_{AB_1})E^{2\alpha}(\rho_{AB_2})\\
&+\left[2^\mu - \mu - \frac{\mu(\mu - 1)}{2} - 1\right]E^\eta(\rho_{AB_2}).
\end{split}
\end{equation}
When \( E^\alpha (\rho_{AB_1}) \leq E^\alpha (\rho_{AB_2}) \), $k=1$, $\mu\geq3$, we have
\begin{equation}\label{15}
\begin{split}
E^\eta(\rho_{A|B_1B_2})\geq & E^\eta(\rho_{AB_2}) + \mu E^{\eta-\alpha}(\rho_{AB_2}) E^\alpha(\rho_{AB_1}) +\frac{\mu(\mu - 1)}{2}E^{\eta - 2\alpha}(\rho_{AB_2})E^{2\alpha}(\rho_{AB_1})\\
&+\left[2^\mu - \mu - \frac{\mu(\mu - 1)}{2}- 1\right]E^\eta(\rho_{AB_1}).
\end{split}
\end{equation}
\qed

Next, we generalize Theorem 1 to $N$-party quantum systems.

\textbf{Theorem 2}. Let $\rho_{AB_1\cdots B_{N-1}} \in \mathcal{H}_A \otimes \mathcal{H}_{B_1} \otimes \cdots \otimes \mathcal{H}_{B_{N-1}}$ be an $N$-party quantum state with $N \geq 4$. Suppose that $E$ is an entanglement measure and that $E^\alpha$ satisfies the monogamy relation. For some integer $m$ satisfying $1 \leq m \leq N-3$, assume that
\[
E^\alpha(\rho_{AB_i}) \geq k \sum_{l=i+1}^{N-1} E^\alpha(\rho_{AB_l}), \qquad i=1,2,\ldots,m,
\]
and
\[
k' E^\alpha(\rho_{AB_j}) \leq \sum_{l=j+1}^{N-1} E^\alpha(\rho_{AB_l}), \qquad j=m+1,\ldots,N-2,
\]
where $k \geq 1$ and $k' \geq 1$. Then, for $\mu=\eta/\alpha $ and $\eta \geq 3\alpha$, we have
\begin{equation}\label{16}
\begin{split}
E^\eta(\rho_{A|B_1\cdots B_{N-1}})
&\geq \sum_{i=1}^{m} \Bigg\{
\left[(k + 1)^\mu - \mu k^{\mu - 1} - \frac{\mu(\mu - 1)}{2}k^{\mu - 2} - k^\mu \right]^{i-1} \\
&\quad \times \Bigg[
E^\eta(\rho_{AB_i})
+ \mu E^{\eta-\alpha}(\rho_{AB_i}) \left( \sum_{l=i+1}^{N-1} E^\alpha(\rho_{AB_l}) \right) \\
&\qquad\quad
+ \frac{\mu(\mu - 1)}{2}E^{\eta-2\alpha}(\rho_{AB_i})
\left( \sum_{l=i+1}^{N-1} E^\alpha(\rho_{AB_l}) \right)^2
\Bigg]
\Bigg\} \\
&\quad + \left[(k + 1)^\mu - \mu k^{\mu - 1} - \frac{\mu(\mu - 1)}{2}k^{\mu - 2} - k^\mu \right]^m
\left[(k'+1)^\mu - k'^\mu \right] \\
&\qquad \times \left[ E^\eta(\rho_{AB_{m+1}}) + \cdots + E^\eta(\rho_{AB_{N-3}}) \right] \\
&\quad + \left[(k + 1)^\mu - \mu k^{\mu - 1} - \frac{\mu(\mu - 1)}{2}k^{\mu - 2} - k^\mu \right]^m \\
&\qquad \times \Bigg\{
\left[(k' + 1)^\mu - \mu k'^{\mu - 1} - \frac{\mu(\mu - 1)}{2}k'^{\mu - 2} - k'^\mu \right]
E^\eta(\rho_{AB_{N-2}}) \\
&\qquad\quad
+ \frac{\mu(\mu - 1)}{2}E^{2\alpha}(\rho_{AB_{N-2}}) E^{\eta-2\alpha}(\rho_{AB_{N-1}}) \\
&\qquad\quad
+ \mu E^\alpha(\rho_{AB_{N-2}}) E^{\eta-\alpha}(\rho_{AB_{N-1}})
+ E^\eta(\rho_{AB_{N-1}})
\Bigg\}.
\end{split}
\end{equation}

\textbf{Proof}. When $E^{\alpha}(\rho_{AB_i}) \geq k \sum_{l=i+1}^{N-1} E^{\alpha}(\rho_{AB_l})$ for $i = 1, 2, \cdots, m$, based on inequalities (\ref{4}) and (\ref{9}), we can obtain
\begin{equation}\label{17}
\begin{split}
E^{\eta}(\rho_{A|B_1B_2\cdots B_{N-1}}) &\geq E^{\eta}(\rho_{AB_1}) + \mu E^{\eta-\alpha}(\rho_{AB_1}) \sum_{l=2}^{N-1} E^{\alpha}(\rho_{AB_l}) \\
&\quad + \frac{\mu(\mu - 1)}{2} E^{\eta-2\alpha}(\rho_{AB_1}) \left( \sum_{l=2}^{N-1} E^{\alpha}(\rho_{AB_l}) \right)^2 \\
&\quad + \left[ (k+1)^\mu - \mu k^{\mu-1} - \frac{\mu(\mu - 1)}{2} k^{\mu-2} - k^\mu \right]
\left( \sum_{l=2}^{N-1} E^{\alpha}(\rho_{AB_l}) \right)^\mu \\
&\geq \sum_{i=1}^{m} \left[ (k+1)^\mu - \mu k^{\mu-1} - \frac{\mu(\mu - 1)}{2} k^{\mu-2} - k^\mu \right]^{i-1}
\Bigg[E^{\eta}(\rho_{AB_i}) \\
&\quad + \mu E^{\eta-\alpha}(\rho_{AB_i}) \sum_{l=i+1}^{N-1} E^{\alpha}(\rho_{AB_l}) \\
&\quad + \frac{\mu(\mu - 1)}{2} E^{\eta-2\alpha}(\rho_{AB_i})
\left( \sum_{l=i+1}^{N-1} E^{\alpha}(\rho_{AB_l}) \right)^2\Bigg] \\
&\quad + \left[ (k+1)^\mu - \mu k^{\mu-1} - \frac{\mu(\mu - 1)}{2} k^{\mu-2} - k^\mu \right]^{m}
\left( \sum_{l=m+1}^{N-1} E^{\alpha}(\rho_{AB_l}) \right)^\mu .
\end{split}
\end{equation}

Next we estimate the last term in (\ref{17}).
When $k' E^{\alpha}(\rho_{AB_j}) \leq \sum_{l=j+1}^{N-1} E^{\alpha}(\rho_{AB_l})$ $(k' \geq 1)$ for \( j = m + 1, \ldots, N - 2 \), inequality (\ref{99}) implies that
$(1+x)^\mu \ge 1 + \big[(k'+1)^\mu - k'^\mu\big]x^\mu.$

By iteratively applying this inequality for \(j=m+1,m+2,\ldots,N-3\), we obtain
\begin{equation}\label{18}
\begin{split}
\left( \sum_{l=m+1}^{N-1} E^\alpha(\rho_{AB_l}) \right)^\mu
&\geq \left( \sum_{l=m+2}^{N-1} E^\alpha(\rho_{AB_l}) \right)^\mu
+ \left[(k'+1)^\mu - k'^\mu\right] E^\eta(\rho_{AB_{m+1}}) \\
&\geq \left[ E^\alpha(\rho_{AB_{N-2}}) + E^\alpha(\rho_{AB_{N-1}}) \right]^\mu \\
&\quad + \left[(k'+1)^\mu - k'^\mu\right]
\left[ E^\eta(\rho_{AB_{m+1}}) +\cdots + E^\eta(\rho_{AB_{N-3}}) \right].
\end{split}
\end{equation}

Finally, taking \(j=N-2\) in the assumption
\(k'E^\alpha(\rho_{AB_{N-2}})\le E^\alpha(\rho_{AB_{N-1}})\),
we again apply inequality (\ref{9}) and obtain
\begin{equation}
\begin{split}
\left[ E^\alpha(\rho_{AB_{N-2}}) + E^\alpha(\rho_{AB_{N-1}}) \right]^\mu
&\geq E^\eta(\rho_{AB_{N-1}})
+ \mu E^\alpha(\rho_{AB_{N-2}})E^{\eta-\alpha}(\rho_{AB_{N-1}}) \\
&\quad + \frac{\mu(\mu-1)}{2}E^{2\alpha}(\rho_{AB_{N-2}})E^{\eta-2\alpha}(\rho_{AB_{N-1}}) \\
&\quad + \left[(k' + 1)^\mu - \mu k'^{\mu - 1} - \frac{\mu(\mu - 1)}{2}k'^{\mu - 2} - k'^\mu\right]
E^\eta(\rho_{AB_{N-2}}).
\end{split}
\end{equation}

Substituting the above estimate into (\ref{18}) and then combining (\ref{17}) with (\ref{18}), we obtain inequality (\ref{16}).
\qed

As a special case, if
\[
E^\alpha(\rho_{AB_i}) \geq \sum_{l=i+1}^{N-1} E^\alpha(\rho_{AB_l}), \qquad i=1,2,\ldots,m,
\]
and
\[
E^\alpha(\rho_{AB_j}) \leq \sum_{l=j+1}^{N-1} E^\alpha(\rho_{AB_l}), \qquad j=m+1,\ldots,N-2,
\]
where $1 \leq m \leq N-3$ and $N \geq 4$, then, for $k=k'=1$ and $\mu \geq 3$, inequality (\ref{16}) reduces to
\begin{equation}\label{19}
\begin{split}
E^\eta(\rho_{A|B_1\cdots B_{N-1}})
&\geq \sum_{i=1}^{m} \Bigg\{
\left[2^\mu - \mu - \frac{\mu(\mu - 1)}{2} - 1 \right]^{i-1} \\
&\quad \times \Bigg[
E^\eta(\rho_{AB_i})
+ \mu E^{\eta-\alpha}(\rho_{AB_i}) \left( \sum_{l=i+1}^{N-1} E^\alpha(\rho_{AB_l}) \right) \\
&\qquad\quad
+ \frac{\mu(\mu - 1)}{2}E^{\eta-2\alpha}(\rho_{AB_i})
\left( \sum_{l=i+1}^{N-1} E^\alpha(\rho_{AB_l}) \right)^2
\Bigg]
\Bigg\} \\
&\quad + \left[2^\mu - \mu - \frac{\mu(\mu - 1)}{2} - 1 \right]^m
\left[2^\mu - 1 \right]
\left[ E^\eta(\rho_{AB_{m+1}}) + \cdots + E^\eta(\rho_{AB_{N-3}}) \right] \\
&\quad + \left[2^\mu - \mu - \frac{\mu(\mu - 1)}{2} - 1 \right]^m \Bigg\{
\left[2^\mu - \mu - \frac{\mu(\mu - 1)}{2} - 1 \right]
E^\eta(\rho_{AB_{N-2}}) \\
&\qquad\quad
+ \frac{\mu(\mu - 1)}{2}E^{2\alpha}(\rho_{AB_{N-2}}) E^{\eta-2\alpha}(\rho_{AB_{N-1}}) \\
&\qquad\quad
+ \mu E^\alpha(\rho_{AB_{N-2}}) E^{\eta-\alpha}(\rho_{AB_{N-1}})
+ E^\eta(\rho_{AB_{N-1}})
\Bigg\}.
\end{split}
\end{equation}

Based on inequality (\ref{9}), we derive the following monogamy relation.

\textbf{Theorem 3}. Let $\rho_{AB_1\cdots B_{N-1}} \in \mathcal{H}_A \otimes \mathcal{H}_{B_1} \otimes \cdots \otimes \mathcal{H}_{B_{N-1}}$ be an $N$-party quantum state. Suppose that $E$ is an entanglement measure and that $E^\alpha$ satisfies the monogamy relation. If
\[
E^{\alpha}(\rho_{AB_i}) \geq k \sum_{l=i+1}^{N-1} E^{\alpha}(\rho_{AB_l}), \qquad i = 1,2,\ldots,N-2,
\]
then
\begin{equation}\label{23}
\begin{split}
E^\eta(\rho_{A|B_1\cdots B_{N-1}})
&\geq \sum_{i=1}^{N-2} \Bigg\{
\left[\left(k+1\right)^\mu - \mu k^{\mu-1} - \frac{\mu(\mu-1)}{2}k^{\mu-2} - k^\mu\right]^{i-1} \\
&\quad \times \Bigg[
E^\eta(\rho_{AB_i})
+ \mu E^{\eta-\alpha}(\rho_{AB_i}) \left( \sum_{j=i+1}^{N-1} E^\alpha(\rho_{AB_j}) \right) \\
&\qquad\quad
+ \frac{\mu(\mu-1)}{2} E^{\eta-2\alpha}(\rho_{AB_i})
\left( \sum_{j=i+1}^{N-1} E^{\alpha}(\rho_{AB_j}) \right)^2
\Bigg]
\Bigg\} \\
&\quad + \left[\left(k+1\right)^\mu - \mu k^{\mu-1} - \frac{\mu(\mu-1)}{2}k^{\mu-2} - k^\mu\right]^{N-2}
E^\eta(\rho_{AB_{N-1}}),
\end{split}
\end{equation}
where $\mu=\eta/\alpha$, $\eta \geq 3\alpha$, and $k \geq 1$.

\textbf{Proof}. For an $N$-party quantum state $\rho_{AB_1\cdots B_{N-1}}$, if
\[
E^{\alpha}(\rho_{AB_i}) \geq k \sum_{j=i+1}^{N-1} E^{\alpha}(\rho_{AB_j}), \qquad i = 1,2,\ldots,N-2,
\]
then, by repeatedly applying inequality (\ref{9}), we obtain
\begin{equation}\label{24}
\begin{split}
E^{\eta}(\rho_{A|B_1\cdots B_{N-1}})
&\geq E^{\eta}(\rho_{AB_1})
+ \mu E^{\eta-\alpha}(\rho_{AB_1}) \sum_{i=2}^{N-1} E^{\alpha}(\rho_{AB_i}) \\
&\quad + \frac{\mu(\mu - 1)}{2} E^{\eta-2\alpha}(\rho_{AB_1})
\left( \sum_{i=2}^{N-1} E^{\alpha}(\rho_{AB_i}) \right)^2 \\
&\quad + \left[ (k+1)^\mu - \mu k^{\mu-1} - \frac{\mu(\mu - 1)}{2} k^{\mu-2} - k^\mu \right]
\left( \sum_{i=2}^{N-1} E^{\alpha}(\rho_{AB_i}) \right)^\mu \\
&\geq \sum_{i=1}^{N-2}
\left[ (k+1)^\mu - \mu k^{\mu-1} - \frac{\mu(\mu - 1)}{2} k^{\mu-2} - k^\mu \right]^{i-1} \\
&\qquad \times \Bigg[
E^{\eta}(\rho_{AB_i})
+ \mu E^{\eta-\alpha}(\rho_{AB_i})
\left( \sum_{j=i+1}^{N-1} E^{\alpha}(\rho_{AB_j}) \right) \\
&\qquad\qquad
+ \frac{\mu(\mu - 1)}{2} E^{\eta-2\alpha}(\rho_{AB_i})
\left( \sum_{j=i+1}^{N-1} E^{\alpha}(\rho_{AB_j}) \right)^2
\Bigg] \\
&\quad + \left[ (k+1)^\mu - \mu k^{\mu-1} - \frac{\mu(\mu - 1)}{2} k^{\mu-2} - k^\mu \right]^{N-2}
E^{\eta}(\rho_{AB_{N-1}}).
\end{split}
\end{equation}
This completes the proof.
\qed

As a special case, if
\[
E^{\alpha}(\rho_{AB_i}) \geq \sum_{l=i+1}^{N-1} E^{\alpha}(\rho_{AB_l}), \qquad i = 1,2,\ldots,N-2,
\]
then, for $k=1$ and $\mu \geq 3$, we have
\begin{equation}\label{25}
\begin{split}
E^{\eta}(\rho_{A|B_1\cdots B_{N-1}})
&\geq \sum_{i=1}^{N-2} \Bigg\{
\left[ 2^\mu - \mu - \frac{\mu(\mu - 1)}{2} - 1 \right]^{i-1} \\
&\quad \times \Bigg[
E^{\eta}(\rho_{AB_i})
+ \mu E^{\eta-\alpha}(\rho_{AB_i})
\left( \sum_{j=i+1}^{N-1} E^{\alpha}(\rho_{AB_j}) \right) \\
&\qquad\quad
+ \frac{\mu(\mu - 1)}{2} E^{\eta-2\alpha}(\rho_{AB_i})
\left( \sum_{j=i+1}^{N-1} E^{\alpha}(\rho_{AB_j}) \right)^2
\Bigg]
\Bigg\} \\
&\quad + \left[ 2^\mu - \mu - \frac{\mu(\mu - 1)}{2} - 1 \right]^{N-2}
E^{\eta}(\rho_{AB_{N-1}}).
\end{split}
\end{equation}

By inequality (\ref{99}), it is straightforward to verify that the lower bounds obtained in Theorems 2 and 3 are tighter than those in [\citealp{37}, Theorems 4 and 5] and [\citealp{33}, Theorems 2 and 3].

\section{Tighter polygamy relation}\label{c}

In this section, we derive a class of tight polygamy relations using entanglement of assistance.

\textbf{Theorem 4}. Let $\rho_{AB_1B_2}\in \mathcal{H}_A \otimes \mathcal{H}_{B_1} \otimes \mathcal{H}_{B_2}$ be any tripartite quantum state. Suppose that $E_a$ is an entanglement of assistance measure and that $E_a^\beta$ satisfies the polygamy relation.

\textit{(i)} For $\gamma = v\beta$ with $0\le v\le 1$ and $k \geq 1$, if
\[
E_a^\beta(\rho_{AB_1}) \geq k E_a^\beta(\rho_{AB_2}),
\]
then
\begin{equation}\label{28}
\begin{split}
E_a^\gamma(\rho_{A|B_1B_2})
\leq\;& E_a^\gamma(\rho_{AB_1})
+ \frac{k v}{k+1} E_a^{\gamma-\beta}(\rho_{AB_1}) E_a^\beta(\rho_{AB_2}) \\
&+\left[ (k+1)^v - \left(\frac{v}{k+1} + 1 \right) k^{v}\right]E_a^\gamma(\rho_{AB_2}).
\end{split}
\end{equation}

\textit{(ii)} For $\gamma = v\beta$ with $0\le v\le 1$ and $k \geq 1$, if
\[
k E_a^\beta(\rho_{AB_1}) \leq E_a^\beta(\rho_{AB_2}),
\]
then
\begin{equation}\label{29}
\begin{split}
E_a^\gamma(\rho_{A|B_1B_2})
\leq\;& E_a^\gamma(\rho_{AB_2})
+ \frac{k v}{k+1} E_a^{\gamma-\beta}(\rho_{AB_2}) E_a^\beta(\rho_{AB_1}) \\
&+\left[ (k+1)^v - \left(\frac{v}{k+1} + 1 \right) k^{v}\right]E_a^\gamma(\rho_{AB_1}).
\end{split}
\end{equation}

\textbf{Proof}. We first prove part \textit{(i)}. Since $E_a^\beta$ satisfies the polygamy relation, it follows from (\ref{5}) that
\[
E_a^\beta(\rho_{A|B_1B_2})
\leq E_a^\beta(\rho_{AB_1}) + E_a^\beta(\rho_{AB_2}).
\]
Hence,
\begin{equation}\label{30}
\begin{split}
E_a^\gamma (\rho_{A|B_1B_2})
&= \left(E_a^\beta(\rho_{A|B_1B_2})\right)^v \\
&\leq \left[ E_a^\beta(\rho_{AB_1}) + E_a^\beta(\rho_{AB_2}) \right]^v \\
&= E_a^\gamma(\rho_{AB_1})
\left( 1 + \frac{E_a^\beta(\rho_{AB_2})}{E_a^\beta(\rho_{AB_1})} \right)^v.
\end{split}
\end{equation}
Since
$E_a^\beta(\rho_{AB_1}) \geq k E_a^\beta(\rho_{AB_2}),$
we have
$0\le \frac{E_a^\beta(\rho_{AB_2})}{E_a^\beta(\rho_{AB_1})}\le \frac{1}{k}.$
Applying Lemma 6 to
$x=\frac{E_a^\beta(\rho_{AB_2})}{E_a^\beta(\rho_{AB_1})},$
we obtain
\begin{equation}
\begin{split}
E_a^\gamma (\rho_{A|B_1B_2})
\leq\;& E_a^\gamma(\rho_{AB_1})
+ \frac{k v}{k+1} E_a^{\gamma-\beta}(\rho_{AB_1}) E_a^\beta(\rho_{AB_2}) \\
&+\left[ (k+1)^v - \left(\frac{v}{k+1} + 1 \right) k^{v}\right]E_a^\gamma(\rho_{AB_2}),
\end{split}
\end{equation}
which proves (\ref{28}).

Part \textit{(ii)} can be proved similarly by interchanging the roles of $B_1$ and $B_2$. \qed

As special cases, when $k=1$ and $0\le v\le 1$, inequality (\ref{28}) reduces to
\begin{equation}\label{31}
\begin{split}
E_a^\gamma(\rho_{A|B_1B_2})
\leq\;& E_a^\gamma(\rho_{AB_1})
+ \frac{v}{2} E_a^{\gamma-\beta}(\rho_{AB_1}) E_a^\beta(\rho_{AB_2}) \\
&+\left( 2^v - \frac{v}{2} - 1\right)E_a^\gamma(\rho_{AB_2}),
\end{split}
\end{equation}
provided that
\[
E_a^\beta(\rho_{AB_1}) \geq E_a^\beta(\rho_{AB_2}).
\]
Similarly, when $k=1$ and $0\le v\le 1$, inequality (\ref{29}) reduces to
\begin{equation}\label{32}
\begin{split}
E_a^\gamma(\rho_{A|B_1B_2})
\leq\;& E_a^\gamma(\rho_{AB_2})
+ \frac{v}{2} E_a^{\gamma-\beta}(\rho_{AB_2}) E_a^\beta(\rho_{AB_1}) \\
&+\left( 2^v - \frac{v}{2} - 1\right)E_a^\gamma(\rho_{AB_1}),
\end{split}
\end{equation}
provided that
\[
E_a^\beta(\rho_{AB_1}) \leq E_a^\beta(\rho_{AB_2}).
\]

Next, we generalize Theorem 4 to $N$-party quantum systems.

\textbf{Theorem 5}. Let $\rho_{AB_1\cdots B_{N-1}} \in \mathcal{H}_A \otimes \mathcal{H}_{B_1} \otimes \cdots \otimes \mathcal{H}_{B_{N-1}}$ be an $N$-party quantum state with $N \ge 4$. Suppose that $E_a$ is an entanglement of assistance measure and that $E_a^\beta$ satisfies the polygamy relation. Let $\gamma=v\beta$ with $0\le v\le 1$. For some integer $m$ satisfying $1\le m\le N-3$, assume that
\[
E_a^\beta(\rho_{AB_i}) \ge k \sum_{l=i+1}^{N-1} E_a^\beta(\rho_{AB_l}),
\qquad i=1,2,\ldots,m,
\]
and
\[
k' E_a^\beta(\rho_{AB_j}) \le \sum_{l=j+1}^{N-1} E_a^\beta(\rho_{AB_l}),
\qquad j=m+1,\ldots,N-2,
\]
where $k\ge 1$ and $k'\ge 1$. Set
\[
\Delta_k(v):=(k+1)^v-\left(1+\frac{v}{k+1}\right)k^v,
\qquad
\Delta_{k'}(v):=(k'+1)^v-\left(1+\frac{v}{k'+1}\right)k'^v.
\]
Then
\begin{equation}\label{33-new}
\begin{split}
E_a^\gamma(\rho_{A|B_1\cdots B_{N-1}})
&\le \sum_{i=1}^{m}\Delta_k(v)^{\,i-1}
\Bigg[
E_a^\gamma(\rho_{AB_i})
+\frac{kv}{k+1}\,
E_a^{\gamma-\beta}(\rho_{AB_i})
\left(\sum_{l=i+1}^{N-1}E_a^\beta(\rho_{AB_l})\right)
\Bigg] \\
&\quad + \Delta_k(v)^m \big[(k'+1)^v-k'^v\big]
\Big[ E_a^\gamma(\rho_{AB_{m+1}})+\cdots+E_a^\gamma(\rho_{AB_{N-3}})\Big] \\
&\quad + \Delta_k(v)^m \Bigg\{
\Delta_{k'}(v)E_a^\gamma(\rho_{AB_{N-2}}) \\
&\qquad\qquad\qquad
+\frac{k'v}{k'+1}\,
E_a^\beta(\rho_{AB_{N-2}})
E_a^{\gamma-\beta}(\rho_{AB_{N-1}})
+E_a^\gamma(\rho_{AB_{N-1}})
\Bigg\}.
\end{split}
\end{equation}

\textbf{Proof}. Let
\[
S_i:=\sum_{l=i}^{N-1}E_a^\beta(\rho_{AB_l}), \qquad i=1,2,\ldots,N-1.
\]
Since $E_a^\beta$ satisfies the polygamy relation,
\[
E_a^\beta(\rho_{A|B_1\cdots B_{N-1}})\le S_1,
\]
and hence
\[
E_a^\gamma(\rho_{A|B_1\cdots B_{N-1}})
=\big(E_a^\beta(\rho_{A|B_1\cdots B_{N-1}})\big)^v
\le S_1^v.
\]

For $i=1,\ldots,m$, the condition
\[
E_a^\beta(\rho_{AB_i})\ge kS_{i+1}
\]
yields
\[
0\le \frac{S_{i+1}}{E_a^\beta(\rho_{AB_i})}\le \frac1k.
\]
Applying Lemma 6 to
\[
S_i^v
=\big(E_a^\beta(\rho_{AB_i})+S_{i+1}\big)^v
=\big(E_a^\beta(\rho_{AB_i})\big)^v
\left(1+\frac{S_{i+1}}{E_a^\beta(\rho_{AB_i})}\right)^v,
\]
we obtain
\[
S_i^v
\le E_a^\gamma(\rho_{AB_i})
+\frac{kv}{k+1}\,
E_a^{\gamma-\beta}(\rho_{AB_i})S_{i+1}
+\Delta_k(v)S_{i+1}^v.
\]
Iterating this inequality for $i=1,\ldots,m$, we get
\begin{equation}\label{eq:T5-mid1}
\begin{split}
E_a^\gamma(\rho_{A|B_1\cdots B_{N-1}})
&\le \sum_{i=1}^{m}\Delta_k(v)^{\,i-1}
\Bigg[
E_a^\gamma(\rho_{AB_i})
+\frac{kv}{k+1}\,
E_a^{\gamma-\beta}(\rho_{AB_i})S_{i+1}
\Bigg] \\
&\quad + \Delta_k(v)^m S_{m+1}^v.
\end{split}
\end{equation}

Next, for $j=m+1,\ldots,N-3$, the assumption
\[
k'E_a^\beta(\rho_{AB_j})\le S_{j+1}
\]
implies
\[
0\le \frac{E_a^\beta(\rho_{AB_j})}{S_{j+1}}\le \frac1{k'}.
\]
Applying inequality \eqref{100}, namely
\[
(1+x)^v\le 1+\big[(k'+1)^v-k'^v\big]x^v,
\qquad 0\le x\le \frac1{k'},
\]
we obtain
\[
S_j^v
=\big(S_{j+1}+E_a^\beta(\rho_{AB_j})\big)^v
\le S_{j+1}^v+\big[(k'+1)^v-k'^v\big]E_a^\gamma(\rho_{AB_j}).
\]
Iterating for $j=m+1,\ldots,N-3$ gives
\begin{equation}\label{eq:T5-mid2}
\begin{split}
S_{m+1}^v
&\le \big[(k'+1)^v-k'^v\big]
\Big[E_a^\gamma(\rho_{AB_{m+1}})+\cdots+E_a^\gamma(\rho_{AB_{N-3}})\Big] \\
&\quad + \big(E_a^\beta(\rho_{AB_{N-2}})+E_a^\beta(\rho_{AB_{N-1}})\big)^v.
\end{split}
\end{equation}

Finally, from
\[
k'E_a^\beta(\rho_{AB_{N-2}})\le E_a^\beta(\rho_{AB_{N-1}}),
\]
we have
\[
0\le \frac{E_a^\beta(\rho_{AB_{N-2}})}{E_a^\beta(\rho_{AB_{N-1}})}\le \frac1{k'}.
\]
Applying Lemma 6 again, we obtain
\begin{equation}\label{eq:T5-mid3}
\begin{split}
&\big(E_a^\beta(\rho_{AB_{N-2}})+E_a^\beta(\rho_{AB_{N-1}})\big)^v \\
&\le E_a^\gamma(\rho_{AB_{N-1}})
+\frac{k'v}{k'+1}\,
E_a^\beta(\rho_{AB_{N-2}})
E_a^{\gamma-\beta}(\rho_{AB_{N-1}})
+\Delta_{k'}(v)E_a^\gamma(\rho_{AB_{N-2}}).
\end{split}
\end{equation}
Substituting \eqref{eq:T5-mid2} and \eqref{eq:T5-mid3} into \eqref{eq:T5-mid1}, and noting that
\[
S_{i+1}=\sum_{l=i+1}^{N-1}E_a^\beta(\rho_{AB_l}),
\]
we arrive at \eqref{33-new}.\qed

Under the assumptions of Theorem 5, if
\[
E_a^\beta(\rho_{AB_i}) \ge \sum_{l=i+1}^{N-1} E_a^\beta(\rho_{AB_l}),
\qquad i=1,2,\ldots,m,
\]
and
\[
E_a^\beta(\rho_{AB_j}) \le \sum_{l=j+1}^{N-1} E_a^\beta(\rho_{AB_l}),
\qquad j=m+1,\ldots,N-2,
\]
then, for $k=k'=1$ and $0\le v\le 1$, \eqref{33-new} reduces to
\begin{equation}\label{36-new-1}
\begin{split}
E_a^\gamma(\rho_{A|B_1\cdots B_{N-1}})
&\le \sum_{i=1}^{m}
\left(2^v-\frac{v}{2}-1\right)^{i-1}
\Bigg[
E_a^\gamma(\rho_{AB_i})
+\frac{v}{2}\,
E_a^{\gamma-\beta}(\rho_{AB_i})
\left(\sum_{l=i+1}^{N-1}E_a^\beta(\rho_{AB_l})\right)
\Bigg] \\
&\quad + \left(2^v-\frac{v}{2}-1\right)^m(2^v-1)
\Big[E_a^\gamma(\rho_{AB_{m+1}})
+\cdots+E_a^\gamma(\rho_{AB_{N-3}})\Big] \\
&\quad + \left(2^v-\frac{v}{2}-1\right)^m
\Bigg[
\left(2^v-\frac{v}{2}-1\right)E_a^\gamma(\rho_{AB_{N-2}}) \\
&\qquad\qquad\qquad
+\frac{v}{2}\,
E_a^\beta(\rho_{AB_{N-2}})
E_a^{\gamma-\beta}(\rho_{AB_{N-1}})
+E_a^\gamma(\rho_{AB_{N-1}})
\Bigg].
\end{split}
\end{equation}

By inequality \eqref{100}, our results are tighter than the corresponding conclusions in [\citealp{33}, Theorems 5 and 6]. The following section further illustrates the tightness of the proposed monogamy and polygamy relations.

\section{Illustrative Examples}

This section is devoted to comparing the monogamy and polygamy relations obtained in this work with existing results. In particular, inequality \eqref{99} implies that the lower bounds given in Theorems 2 and 3 are tighter than those in [\citealp{37}, Theorems 4 and 5] and [\citealp{33}, Theorems 2 and 3], whereas inequality \eqref{100} shows that our polygamy relation improves upon the corresponding results in [\citealp{33}, Theorems 5 and 6]. To further illustrate the tightness of the derived inequalities, we present the following example. Here, concurrence is chosen as the bipartite entanglement measure for the monogamy relation, while concurrence of assistance is used for the polygamy relation. The example below shows that our bounds are tighter than those reported in Refs.~\cite{33,37}.

\textbf{Example 1}. Under local unitary transformations, any three-qubit pure state can be written as
\begin{equation}\label{37}
\begin{split}
|\varphi_{AB_1B_2}\rangle
=\lambda_0|000\rangle+\lambda_1 e^{i\phi}|100\rangle+\lambda_2|101\rangle+\lambda_3|110\rangle+\lambda_4|111\rangle,
\end{split}
\end{equation}
where $0\leq \phi \leq \pi$, $\lambda_i\geq 0$ for $i=0,1,2,3,4$, and $\sum_{i=0}^4 \lambda_i^2=1$.

We first consider the choice
\[
\lambda_0=\frac{\sqrt{2}}{3},\qquad
\lambda_1=\lambda_4=0,\qquad
\lambda_2=\frac{\sqrt{2}}{3},\qquad
\lambda_3=\frac{\sqrt{5}}{3}.
\]
Then
\[
C(\rho_{AB_1})=\frac{2\sqrt{10}}{9},\qquad
C(\rho_{AB_2})=\frac{4}{9},\qquad
C(\rho_{A|B_1B_2})=\frac{2\sqrt{14}}{9}.
\]

By the result of \cite{33}, one has
\begin{equation}\label{38}
\begin{split}
C^{2\mu}(|\varphi_{A|B_1B_2}\rangle)
\geq \left(\frac{40}{81}\right)^\mu
+\frac{k\mu}{k+1}\left(\frac{40}{81}\right)^{\mu-1}\frac{16}{81}
+\left[(k+1)^\mu-\left(1+\frac{\mu}{k+1}\right)k^\mu\right]\left(\frac{16}{81}\right)^\mu.
\end{split}
\end{equation}

From \cite{37}, it follows that
\begin{equation}\label{39}
\begin{split}
C^{2\mu}(|\varphi_{A|B_1B_2}\rangle)
\geq \left(\frac{40}{81}\right)^\mu
+\mu\left(\frac{40}{81}\right)^{\mu-1}\frac{16}{81}
+\left[(k+1)^\mu-\mu k^{\mu-1}-k^\mu\right]\left(\frac{16}{81}\right)^\mu.
\end{split}
\end{equation}

By inequality \eqref{11}, we obtain
\begin{equation}\label{40}
\begin{split}
C^{2\mu}(|\varphi_{A|B_1B_2}\rangle)
\geq & \left(\frac{40}{81}\right)^\mu
+\mu\left(\frac{40}{81}\right)^{\mu-1}\frac{16}{81}
+\frac{\mu(\mu-1)}{2}\left(\frac{40}{81}\right)^{\mu-2}\left(\frac{16}{81}\right)^2 \\
&+\left[(k+1)^\mu-\mu k^{\mu-1}-\frac{\mu(\mu-1)}{2}k^{\mu-2}-k^\mu\right]\left(\frac{16}{81}\right)^\mu.
\end{split}
\end{equation}

\noindent\makebox[\textwidth]{%
  \includegraphics[width=0.75\textwidth]{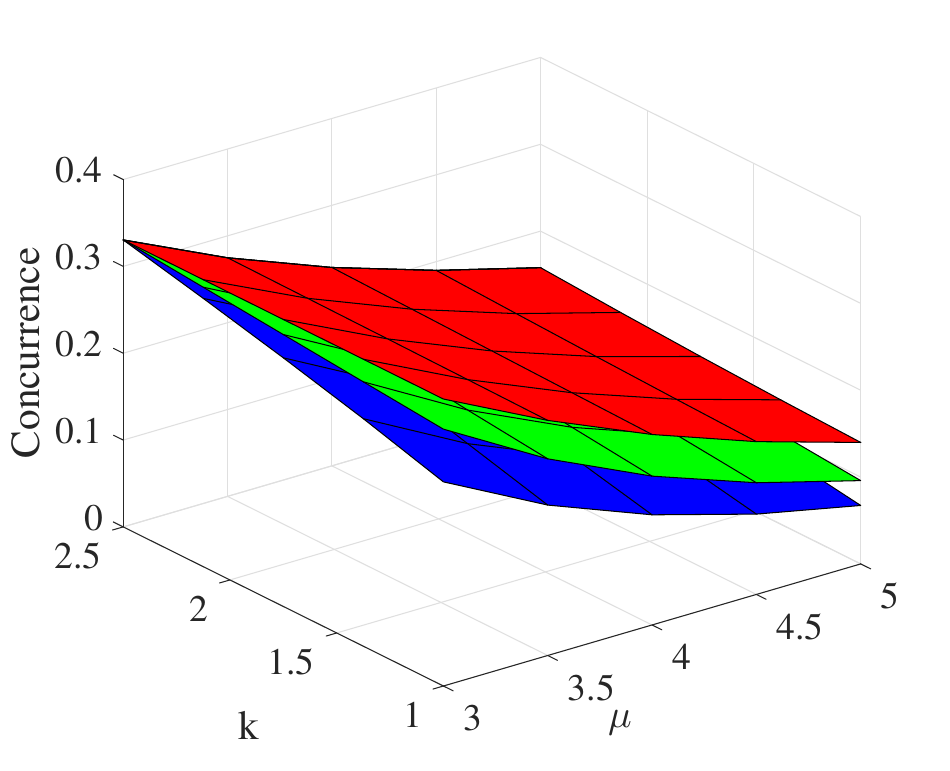}%
}
\vspace{0.5em}

\textbf{Fig. 1.} The red, green, and blue surfaces correspond to the lower bounds given by inequalities \eqref{40}, \eqref{39}, and \eqref{38}, respectively.

\vspace{1em}

Figure 1 shows that inequality \eqref{40} provides a tighter lower bound than those in \eqref{38} and \eqref{39}.

For the polygamy relation, we take
\[
\lambda_0=\frac{1}{2},\qquad
\lambda_1=\lambda_4=\frac{\sqrt{2}}{12},\qquad
\lambda_2=\frac{\sqrt{2}}{2},\qquad
\lambda_3=\frac{\sqrt{2}}{3}.
\]
Then
\[
C_a(\rho_{AB_1})=\frac{\sqrt{34}}{12},\qquad
C_a(\rho_{AB_2})=\frac{\sqrt{74}}{12},\qquad
C_a(\rho_{A|B_1B_2})=\frac{\sqrt{106}}{12}.
\]

According to [\citealp{33}, Theorem 5], one has
\begin{equation}\label{41}
\begin{split}
C_a^{2v}(\rho_{A|B_1B_2})
\leq & \left(\frac{37}{72}\right)^v
+\frac{k^2v}{(k+1)^2}\left(\frac{37}{72}\right)^{v-1}\frac{17}{72} \\
&+\left[(k+1)^v-\left(\frac{kv}{(k+1)^2}+1\right)k^v\right]\left(\frac{17}{72}\right)^v.
\end{split}
\end{equation}

On the other hand, inequality \eqref{28} yields
\begin{equation}\label{42}
\begin{split}
C_a^{2v}(\rho_{A|B_1B_2})
\leq & \left(\frac{37}{72}\right)^v
+\frac{kv}{k+1}\left(\frac{37}{72}\right)^{v-1}\frac{17}{72} \\
&+\left[(k+1)^v-\left(\frac{v}{k+1}+1\right)k^v\right]\left(\frac{17}{72}\right)^v.
\end{split}
\end{equation}

\noindent\makebox[\textwidth]{%
  \includegraphics[width=0.75\textwidth]{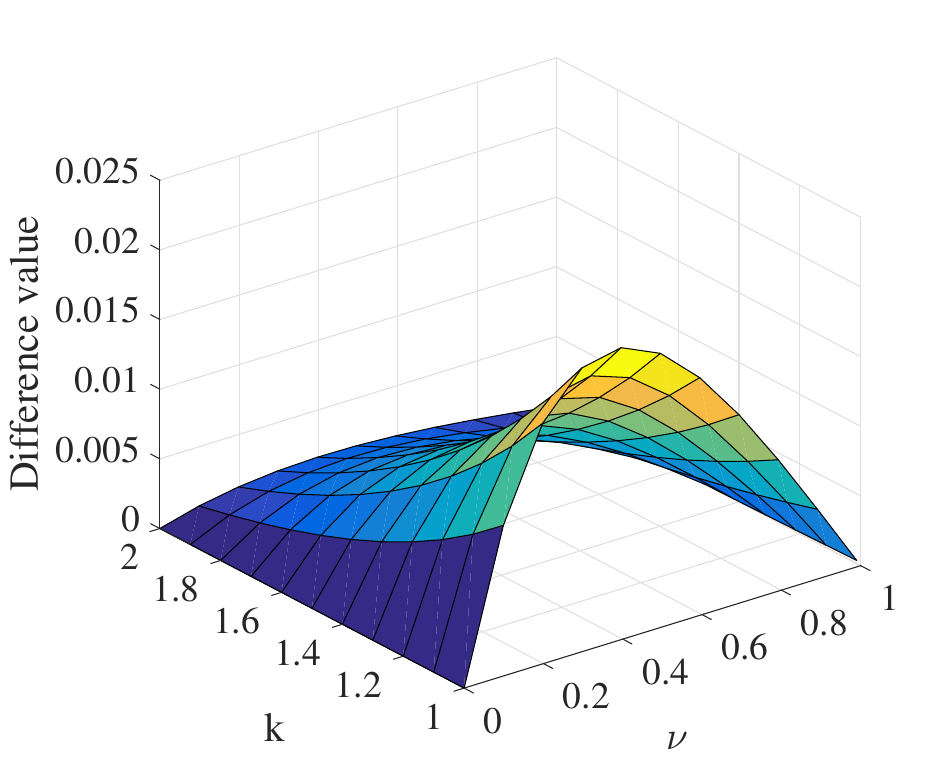}%
}
\vspace{0.5em}

\textbf{Fig. 2.} The figure shows the difference between the right-hand sides of inequalities \eqref{41} and \eqref{42}.

\vspace{1em}

Figure 2 further confirms that inequality \eqref{42} gives a tighter upper bound than \eqref{41}.

\section{Conclusion}

In this work, we have established a class of tighter monogamy and polygamy relations for multipartite quantum entanglement. By introducing a new inequality and providing a rigorous proof of its validity, we derived improved monogamy inequalities for the $\alpha$th power of a bipartite entanglement measure $E$ that satisfies the monogamy relation. These results were further generalized from tripartite systems to general multipartite quantum systems under certain constraint conditions. By selecting appropriate parameters and comparing the resulting bounds with those reported in Refs.~\cite{33,37}, we showed that the obtained monogamy relations provide tighter lower bounds than the existing ones.

We also investigated the corresponding polygamy relations in terms of the $\beta$th power of the entanglement of assistance. Based on the newly established inequalities and the definition of entanglement polygamy, we derived a family of refined polygamy inequalities and extended them from tripartite systems to general $N$-partite quantum systems. Numerical comparisons with the existing result in Ref.~\cite{33} demonstrate that the upper bounds derived in this work are tighter, providing a more accurate characterization of entanglement distribution.

These results provide a refined description of entanglement distribution in multipartite quantum systems and may be useful for further studies on the structure and shareability of quantum entanglement.

\section*{ACKNOWLEDGMENTS}

This work was supported by the National Pre-research Funds of Hebei GEO University under Grant No.~KY2025YB15.

\end{document}